\documentclass[preprint,onecolumn,showkeys,amsmath,amssymb,aps,prb,floatfix]{revtex4-1}
\usepackage{graphicx}
\usepackage{dcolumn}
\usepackage{bm}
\usepackage{hyperref}

	\usepackage{fancyhdr}
\renewcommand{\thispagestyle}[1]{} 

\begin{document}

\preprint{}

\title{Normal and inverse magnetocaloric effect in magnetic multilayers with antiferromagnetic interlayer coupling}

\author{Karol Sza{\l}owski}
 \email{kszalowski@uni.lodz.pl}
\author{Tadeusz Balcerzak}%
\affiliation{%
Department of Solid State Physics, Faculty of Physics and Applied Informatics, University of
{\L}\'{o}d\'{z},\\ ulica Pomorska 149/153, 90-236 {\L}\'{o}d\'{z}, Poland
}%

\date{\today}

\begin{abstract}
The thermodynamics of a spin-1/2 magnetic multilayer system with antiferromagnetic interplanar couplings is studied using the Pair Approximation method. The special attention is paid to magnetocaloric properties, quantified by isothermal entropy change. The multilayer consists of two kinds of magnetic planes, one of which is diluted. The intraplanar couplings in both planes have arbitrary anisotropy ranging between Ising and isotropic Heisenberg interactions. The phase diagram related to the occurrence of magnetic compensation phenomenon is constructed and discussed. Then the isothermal entropy change is discussed as a function of interaction parameters, magnetic component concentration and external magnetic field amplitude. The ranges of normal and inverse magnetocaloric effect are found and related to the presence or absence of compensation.
\end{abstract}

\pacs{Valid PACS appear here}
\maketitle


\section{Introduction}

Magnetocaloric effect (MCE) belongs to particularly intensely investigated subjects within physics of magnetic systems\cite{mcalbook,Tishin2014,Gschneidner2005,Szymczak}, mainly owing to its high potential for applications in modern and highly expected solid-state refrigeration and for construction of energy-saving devices. The works focus on extensive searching for novel materials exhibiting better performance \cite{Sandeman2012,Manosa2013}. In parallel, there is a constant need for theoretical models giving physical insight into the phenomenon and allowing to describe the experimental results. The recently used analytical approaches to the theory of MCE usually exploit scaling relations and employ scaling-based equations of state, making use of mean field approximation for constructing the thermodynamics of magnetic systems \cite{Pelka2013,Franco2008,Franco2010,Franco2012,AmaralInTech,Amaral2007,deOliveira2010,deOliveira2014,Dong2008,Basso2014}. Such approaches are very successful from the practical point 
of view, enabling for example extrapolation of the experimental results and allowing for construction of universal dimensionless functions \cite{Amaral2007,Franco2008,Franco2012,AmaralInTech,Pelka2013}. However, they are not free from some empirical parameters. On the other hand, also Monte Carlo simulations are used to predict theoretically the magnetocaloric properties of materials \cite{Nobrega2005,Nobrega2006,Nobrega2007,Nobrega2011,Buchelnikov2011,Singh2013}.
Let us also mention the existence of a number of exactly solved spin models for which magnetocaloric quantities have been discussed within various approaches, such as Jordan-Wigner transformation \cite{Zhitomirsky2004,Topilko2012}, Bethe ansatz-based quantum transfer matrix and non-linear integral equations method \cite{Trippe2010,Ribeiro2010}, or generalized classical transfer-matrix method and decoration-iteration mapping \cite{Canova2006,Pereira2009,Strecka2014,Ohanyan2012,Canova2014}.

Among various systems exhibiting MCE, ferrimagnets draw considerable attention. In such magnets, an inverse MCE can occur, consisting in decrease of the entropy when the external magnetic field changes isothermally from non-zero to zero value, whereas for the normal MCE the entropy rises under such conditions. This phenomenon is both observed experimentally (e.g. \cite{Burzo2010,Zhang2010,Reis2010}), 
for instance, in the following materials: $\rm MnBr4H_2O$, $\rm Yb_3Fe_5O_{12}$, $\rm Ni_{50}Mn_{34}In_{16}$, $\rm CoMnSi$, $\rm Mn_{1.82}V_{0.18}Sb$,
and its existence has been qualitatively explained \cite{vonRanke2009a,vonRanke2009b,vonRanke2010,Biswas2013a,Biswas2013b,Alho2010} and predicted for some classes of non-trivial magnets (like low-dimensional \cite{Qi2012} and frustrated systems \cite{Zukovic2013}). On the other hand, one of the directions in studies of MCE is turning the attention to magnetic multilayers \cite{Caballero2012,Florez2013}. In particular, layered magnets with antiferromagnetic coupling between the layers can exhibit ferrimagnetic ground state. The temperature dependence of total magnetization of such systems may show the presence of 
magnetic compensation, which has been extensively studied in various classes of layered and similar magnets \cite{Kaneyoshi1993,Kaneyoshi1995,Kaneyoshi1996,Kaneyoshi2011,Kaneyoshi2012a,Kaneyoshi2012b,Kaneyoshi2013a,
Kaneyoshi2013b,Kaneyoshi2013c,Kantar2014,Jascur1997a,Jascur1997b,Yuksel2013,Bobak2011,Oitmaa2003,Oitmaa2005,Veiller2000,Balcerzak2014}. Also, MCE in vicinity of the compensation temperature has been a subject of studies \cite{Ma2013}. We mention that the coexistence of normal and inverse MCE is in general possible in a class of magnets exhibiting rich structure of the ground-state phase diagram, in particular in Ising-Heisenberg chains \cite{Zhitomirsky2004,Topilko2012,Trippe2010,Ribeiro2010,Canova2006,Pereira2009,Strecka2014,Ohanyan2012,Canova2014} or even some zero-dimensional structures \cite{Cisarova2014}. However, we consider the multilayer geometry to be of particular interest.

Motivated by the findings connected with normal and inverse MCE in layered magnets, we present a theoretical study of a magnetic spin-1/2 multilayer with antiferromagnetic interlayer couplings. 
We assume that the planes A and B forming multilayer are magnetically non-equivalent, having different exchange parameters and different anisotropy. Moreover, one of the plane (that one which is magnetically stronger) is randomly diluted. In such a system the compensation phenomenon can occur, whose characteristic compensation temperature can be much lower than the N\'eel temperature, and can be modified by the degree of dilution \cite{Balcerzak2014}. Since in the vicinity of compensation temperature the inverse MCE can be expected, the temperature range  of occurrence of this effect (and its strength) could be, to some extent, controlled by the dilution parameter, even though the other material parameters (exchange integrals) remain constant. Apart from pure theoretical interest in the model we think that such a possibility may also inspire the researchers for its experimental realization.\\

The paper develops a theoretical model and its thermodynamic description. Then the conditions for presence of magnetic compensation phenomenon are discussed and the magnetocaloric properties (the isothermal entropy change and the adiabatic cooling rate) are analysed. The numerical results are extensively illustrated in plots.

\section{Theoretical model}

\begin{figure}
\includegraphics[scale=0.25]{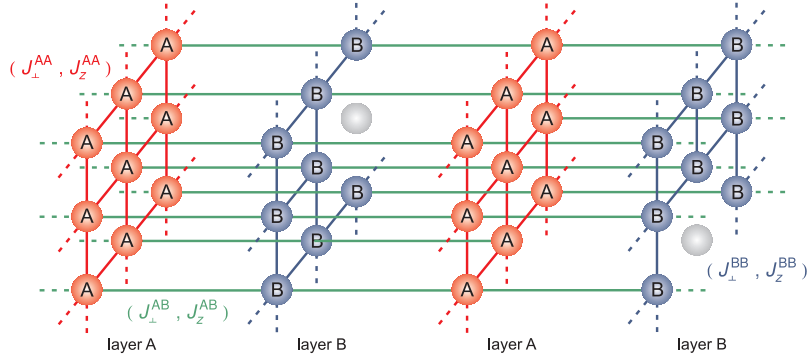}
\caption{\label{fig:fig1}A schematic view of the multilayer consisting of two inequivalent kinds of planes, $A$ and $B$. The intraplanar couplings are $J^{AA}_{x}=J^{AA}_{y}=J^{AA}_{\perp}$, $J^{AA}_{z}$ and $J^{BB}_{x}=J^{BB}_{y}=J^{BB}_{\perp}$, $J^{BB}_{z}$, respectively. The interplanar coupling is $J^{AB}_{z}$. The plane $B$ is randomly diluted. }
\end{figure}

The subject of our interest is a magnetic multilayer, in which the spins are located at the sites of simple cubic (sc) crystalline lattice. The system in question is composed of inequivalent parallel monolayers which are stacked alternately and are further called A and B planes. Each single plane is therefore a simple quadratic lattice. All sites of plane A are populated by magnetic atoms, while plane B is site-diluted, so that the concentration of magnetic atoms equals there to $p$. All magnetic atoms are assumed to have spin 1/2. The schematic view of the multilayer is presented in Fig.~\ref{fig:fig1}.

The Hamiltonian of the system takes the following form:
\begin{eqnarray}
\mathcal{H}&=&-\sum_{\left\langle i\in A,j\in A \right\rangle}^{}{\left[J_{\perp}^{AA}\left(S_{x}^{i}S_{x}^{j}+S_{y}^{i}S_{y}^{j}\right)+J_{z}^{AA}S_{z}^{i}S_{z}^{j}\right]}-\sum_{\left\langle i\in A,j\in B \right\rangle}^{}{J_{z}^{AB}S_{z}^{i}S_{z}^{j}\xi_j}\nonumber\\
&&-\sum_{\left\langle i\in B,j\in B \right\rangle}^{}{\left[J_{\perp}^{BB}\left(S_{x}^{i}S_{x}^{j}+S_{y}^{i}S_{y}^{j}\right)+J_{z}^{BB}S_{z}^{i}S_{z}^{j}\right]\xi_i\xi_j}-h\sum_{i\in A}^{}{S^{i}_{z}}-h\sum_{j\in B}^{}{S^{j}_{z}\xi_j}.
\label{eq1}
\end{eqnarray}
The exchange integrals characterizing the interactions between various nearest-neighbour (NN) spins are indicated schematically in Fig.~\ref{fig:fig1}. In particular, $J^{AA}_{x}=J^{AA}_{y}=J^{AA}_{\perp}$ and $J^{AA}_{z}$ are intraplanar couplings in plane A, while $J^{BB}_{x}=J^{BB}_{y}=J^{BB}_{\perp}$ and $J^{BB}_{z}$ are intraplanar couplings in plane B. All the intraplanar interactions are ferromagnetic. Spins in different neighbouring planes interact antiferromagnetically, and the corresponding exchange integral is $J^{AB}_{z}<0$. External magnetic field is denoted by $h$. 

In our study we consider the interplanar couplings in planes A and B exhibiting easy axis-type anisotropy, i.e. $0\leq J^{\gamma \gamma}_{\perp} \le J^{\gamma \gamma}_{z}$ for $\gamma=A,B$. This corresponds to the limiting cases of all-Ising interactions (when $J^{\gamma \gamma}_{\perp}=0$) and the isotropic Heisenberg couplings (with $J^{\gamma \gamma}_{\perp}=J^{\gamma \gamma}_{z}$).  The interaction between the planes is taken in Ising form, since $J^{AB}_{\perp}=0$ might lead to some non-classical magnetic ground state of the system. The site occupation operators $\xi_{j}$ take the values of $\xi_j=0,1$ and are introduced to describe random site dilution in the plane $B$ whereas the concentration of magnetic atoms in plane B equals to their configurational average, $\left< \xi_j \right> =p$.

In order to describe the thermodynamic properties of the model in question, we use the Pair Approximation (PA), which method has been employed by us for characterization of various spin-1/2 magnetic systems \cite{Balcerzak2009a,Balcerzak2009b,Szalowski2011,Szalowski2012,Szalowski2013,Balcerzak2014}, including the magnetocaloric studies \cite{Szalowski2011,Szalowski2013} and studies of layered magnets \cite{Szalowski2012,Szalowski2013,Balcerzak2014}. The method allows to find a Gibbs free energy within fully self-consistent approach, which allows for further determination of all interesting magnetic properties by employing thermodynamic identities.

In order to obtain the Gibbs energy we employ the general relationship:
\begin{equation}
G=\left<\mathcal{H}\right> - ST
\label{neweq1}
\end{equation}
where $\left<\mathcal{H}\right>$ is the average value of Hamiltonian (\ref{eq1}) containing external field (i.e., enthalpy of the system), and $S$ is the total entropy. The entropy is obtained by the cumulant technique where only single-site and pair entropy cumulants are taken into account:
\begin{equation}
S=\frac{N}{2}\left(\sigma^A+p\sigma^B\right)+N\left[\left(\sigma^{AA}-2\sigma^A\right)+p^2\left(\sigma^{BB}-2\sigma^B\right)+p\left(\sigma^{AB}-\sigma^A-\sigma^B\right)\right].
\label{neweq2}
\end{equation}
In Eq.~(\ref{neweq2}),  $N$ is the total number of lattice sites, whereas $\sigma^\gamma$ and $\sigma^{\gamma \delta}$ ($\gamma=A,B$; $\delta=A,B$) are single-site and pair entropies, respectively. Eq.~(\ref{neweq2}) can be re-written in a more convenient form presenting the entropy per lattice site:
\begin{equation}
\frac{S}{N}=
\sigma^{AA}+p\sigma^{AB}+p^2\sigma^{BB}-\left(\frac{3}{2}+p\right)\sigma^A-p\left(\frac{1}{2}+2p\right)\sigma^B.
\label{neweq3}
\end{equation}
The single-site and pair entropies are defined by the expressions:
\begin{equation}
\sigma^\gamma=-k_{\rm B}\left<\ln \rho^{\gamma}_{i}\right> 
\label{neweq4}
\end{equation}
and
\begin{equation}
\sigma^{\gamma \delta}=-k_{\rm B}\left<\ln \rho^{\gamma \delta}_{i j}\right> ,
\label{neweq5}
\end{equation}
where $\rho^{\gamma}_{i}$ and $\rho^{\gamma \delta}_{i j}$ are single-site and pair density matrices, respectively. These matrices are of the form:
\begin{equation}
\rho^{\gamma}_{i}=e^{\beta\left(G^{\gamma} -\mathcal{H}^{\gamma}_{i}\right)}
\label{neweq6}
\end{equation}
and
\begin{equation}
\rho^{\gamma \delta}_{i j}=e^{\beta\left(G^{\gamma \delta} -\mathcal{H}^{\gamma \delta}_{i j}\right)}
\label{neweq7}
\end{equation}
($\beta=1/k_{\rm B}T$). \\
In Eqs.~(\ref{neweq6}) and (\ref{neweq7}), $\mathcal{H}^{\gamma}_{i}$ and $\mathcal{H}^{\gamma \delta}_{i j}$ denote single-site ($i \epsilon \gamma$) and NN-pair ($i\epsilon \gamma;  j\epsilon \delta$) cluster Hamiltonians \cite{Szalowski2012} whereas $G^{\gamma}$ and $G^{\gamma \delta}$ are corresponding cluster Gibbs energies. With the help of expressions (\ref{neweq4}) and (\ref{neweq5}) the cluster entropies can also be expressed as:
\begin{equation}
-\sigma^{\gamma}T=G^{\gamma} -\left<\mathcal{H}^{\gamma}_{i}\right> 
\label{neweq8}
\end{equation}
and
\begin{equation}
-\sigma^{\gamma \delta}T=G^{\gamma \delta} -\left<\mathcal{H}^{\gamma \delta}_{i j}\right>. 
\label{neweq9}
\end{equation}
Substituting Eqs.~(\ref{neweq8}) and (\ref{neweq9}) into the expression for the total entropy (\ref{neweq3}), the total Gibbs energy per lattice site is finally expressed from Eq.~(\ref{neweq1}) in terms of single-site and NN-pair Gibbs energies as:
\begin{equation}
\frac{G}{N}=G^{AA}+pG^{AB}+p^{2}G^{BB}-\left(\frac{3}{2}+p\right)G^{A}-p\left(\frac{1}{2}+2p\right)G^{B},
\label{eq2}
\end{equation}
where the terms representing cluster enthalpies $\left< \mathcal{H}^{\gamma}_{i}\right>$ and $\left<\mathcal{H}^{\gamma \delta}_{i j}\right>$ have been cancelled in Eq.~(\ref{neweq1})  with the total enthalpy term $\left<\mathcal{H}\right>$.
The Gibbs energies are then obtained from normalization condition for density matrices (\ref{neweq6}) and (\ref{neweq7}) and are given by \cite{Szalowski2012}:
\begin{equation}
G^{\gamma}=-k_{\rm B}T \,\ln\left\{2\cosh\left[\frac{\beta}{2}\left(\Lambda^{\gamma}+h\right)\right]\right\} 
\label{eq3}
\end{equation}
for single-site clusters ($\gamma =A,B$), while for NN-pairs the Gibbs energies are of the form:
\begin{equation}
G^{\gamma \delta}=-k_{\rm B}T \,\ln\left\{2\exp\left(\frac{\beta J^{\gamma \delta}_{z}}{4}\right)\cosh\left[\beta\left(\Lambda^{\gamma \delta}+h\right)\right]+2\exp\left(-\frac{\beta J^{\gamma \delta}_{z}}{4}\right)\cosh\left[\frac{\beta}{2}\sqrt{\left(\Delta^{\gamma \delta}\right)^2+\left(J_{\perp}^{\gamma \delta}\right)^2}\,\right]\right\} 
\label{eq4},
\end{equation}
where $\gamma=A,B$ and $\delta=A,B$.\\
The above equations contain the parameters $\Lambda^{\gamma}$, $\Lambda^{\gamma \delta}$ and $\Delta^{\gamma \delta}$, which can be expressed by four independent variational parameters $\lambda^{AA}$, $\lambda^{BB}$, $\lambda^{AB}$ and $\lambda^{BA}$ in the following way:
\begin{eqnarray}
\Lambda^{A}&=&4\lambda^{AA}+2p\lambda^{AB}\nonumber\\
\Lambda^{B}&=&4p\lambda^{BB}+2\lambda^{BA}\nonumber\\
\Lambda^{AA}&=&3\lambda^{AA}+2p\lambda^{AB}\nonumber\\
\Lambda^{BB}&=&\left(4p-1\right)\lambda^{BB}+2\lambda^{BA}\nonumber\\
\Lambda^{AB}&=&2\left(\lambda^{AA}+p\lambda^{BB}\right)+\frac{1}{2}\left(2p-1\right)\lambda^{AB}+\frac{1}{2}\lambda^{BA}\nonumber\\
\Delta^{AA}&=&0 \nonumber\\
\Delta^{BB}&=&0 \nonumber\\
\Delta^{AB}&=&4\left(\lambda^{AA}-p\lambda^{BB}\right)+\left(2p-1\right)\lambda^{AB}-\lambda^{BA}
\label{eq5}.
\end{eqnarray}
The parameters are found from the variational principle for the Gibbs energy minimization, $\partial G/\partial \lambda^{\gamma \delta}=0$. As a result, the following set of four self-consistent equations for parameters $\lambda^{\gamma \delta}$ is obtained:
\begin{eqnarray}
\tanh\left[\frac{\beta}{2}\left(\Lambda^{A}+h\right)\right]&=&\frac{\exp\left(\frac{\beta J^{AA}_{z}}{4}\right)\sinh\left[\beta\left(\Lambda^{AA}+h\right)\right]}{\exp\left(\frac{\beta J^{AA}_{z}}{4}\right)\cosh\left[\beta\left(\Lambda^{AA}+h\right)\right]+\exp\left(-\frac{\beta J^{AA}_{z}}{4}\right)\cosh\left(\frac{\beta J^{AA}_{\perp}}{2}\right)}
\label{eq7}
\\
\tanh\left[\frac{\beta}{2}\left(\Lambda^{A}+h\right)\right]&=&\frac{\exp\left(\frac{\beta J^{AB}_{z}}{4}\right)\sinh\left[\beta\left(\Lambda^{AB}+h\right)\right]+ \exp\left(-\frac{\beta J^{AB}_{z}}{4}\right)\sinh\left(\frac{\beta \Delta^{AB}}{2}\right)}{\exp\left(\frac{\beta J^{AB}_{z}}{4}\right)\cosh\left[\beta\left(\Lambda^{AB}+h\right)\right]+\exp\left(-\frac{\beta J^{AB}_{z}}{4}\right)\cosh\left(\frac{\beta \Delta^{AB}}{2}\right)}
\label{eq8}
\\
\tanh\left[\frac{\beta}{2}\left(\Lambda^{B}+h\right)\right]&=&\frac{\exp\left(\frac{\beta J^{BB}_{z}}{4}\right)\sinh\left[\beta\left(\Lambda^{BB}+h\right)\right]}{\exp\left(\frac{\beta J^{BB}_{z}}{4}\right)\cosh\left[\beta\left(\Lambda^{BB}+h\right)\right]+\exp\left(-\frac{\beta J^{BB}_{z}}{4}\right)\cosh\left(\frac{\beta J^{BB}_{\perp}}{2}\right)}
\label{eq9}
\\
\tanh\left[\frac{\beta}{2}\left(\Lambda^{B}+h\right)\right]&=&\frac{\exp\left(\frac{\beta J^{AB}_{z}}{4}\right)\sinh\left[\beta\left(\Lambda^{AB}+h\right)\right]- \exp\left(-\frac{\beta J^{AB}_{z}}{4}\right)\sinh\left(\frac{\beta \Delta^{AB}}{2}\right)}{\exp\left(\frac{\beta J^{AB}_{z}}{4}\right)\cosh\left[\beta\left(\Lambda^{AB}+h\right)\right]+\exp\left(-\frac{\beta J^{AB}_{z}}{4}\right)\cosh\left(\frac{\beta \Delta^{AB}}{2}\right)}
\label{eq10}.
\end{eqnarray}

The above set of non-linear, self-consistent equations can be solved only numerically and for this purpose in the present work we used Mathematica software package \cite{Mathematica}.

The Gibbs free energy (Eq.~\ref{eq2}) is in principle the function of the external field $h$ and temperature $T$. The self-consistency of the thermodynamic description yields the possibility of determining all the other interesting thermodynamic quantities form the appropriate identities. For instance, the total magnetization per site can be found from:
\begin{equation}
m_{tot}=-\left(\frac{\partial G}{\partial h}\right)_{T},
\label{eq11}
\end{equation}
leading finally to:
\begin{equation}
m_{tot}=\frac{1}{2}\left(m_A+pm_B\right),
\label{eq12}
\end{equation}
where 
\begin{equation}
m_{\gamma}=\frac{1}{2}\tanh\left[\frac{\beta}{2}\left(\Lambda^{\gamma}+h\right)\right] 
\label{eq13}
\end{equation}
for $\gamma = A,B$. $m_A$ and $m_B$ are the magnetizations per occupied lattice site in plane A and B, respectively. 

For the antiferromagnetic interaction $J^{AB}_{z}$ between the planes $A$ and $B$, magnetizations $m_A$ and $m_B$ have opposite signs. As the magnetizations vary with the temperature, a compensation phenomenon can take place, which means that $m_{tot}=m_A+pm_B=0$ for $m_{A},m_{B}\neq 0$ and $T<T_{c}$, where $T_{c}$ is the phase transition temperature. Let us mention that the occurrence of such a phenomenon in magnetic bilayer has been a subject of our extensive study in Ref.~\cite{Balcerzak2014}. The conditions for its existence in the present system, i.e. the magnetic multilayer, will be discussed in further part of the paper. We also mention that the critical temperature for layered ferro- and ferrimagnets with
anisotropic interactions has been a subject of numerous recent studies (e.g. \cite{Szalowski2012,Balcerzak2014,Akinci}).

The magnetocaloric effect can be characterized mostly with the help of such quantity as the entropy change $\Delta S_{T}$ in the process of isothermal demagnetization between the external field $h>0$ and $h=0$ \cite{mcalbook,Szalowski2011}. It can be expressed as $\Delta S_{T}=S\left(T,h\right)-S\left(T,h=0\right)$. The magnetic entropy of the system is found from the identity $S=-\left(\partial G/\partial T\right)_{h}$. In the presented convention, positive value of $\Delta S_{T}$ corresponds to normal MCE, while the negative value denotes inverse MCE. Another magnetocaloric quantity of interest is the temperature change vs. magnetic field under adiabatic (isentropic) process, usually characterized with the help of cooling rate $\Gamma_{S}=\left(\partial T/\partial h\right)_{S}=-\left(\partial m/\partial T\right)_{h}/\left(\partial S/\partial T\right)_{h}$ \cite{mcalbook,Szalowski2011}.

\section{Numerical results and discussion}

\begin{figure}
\includegraphics[scale=0.25]{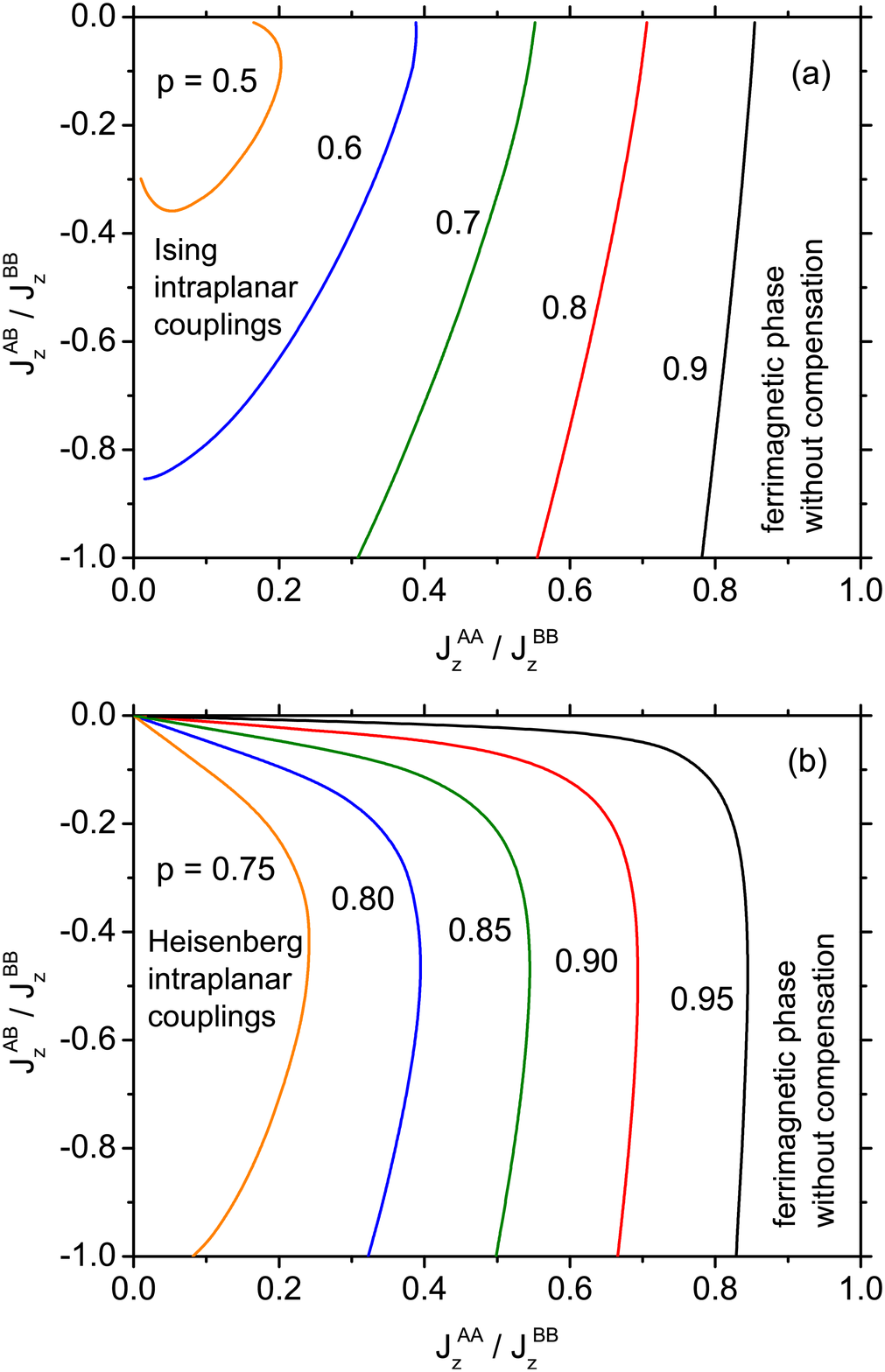}
\caption{\label{fig:fig2} Phase diagram for the multilayer ferrimagnetic system showing the phase boundaries between the ranges of parameters in which the magnetic compensation phenomenon is present and absent. The ferrimagnetic phase without compensation is the phase right of the boundary line. Various concentrations of magnetic component in plane B are considered. The intraplanar couplings are of Ising type (a) and isotropic Heisenberg type (b). }
\end{figure}

\begin{figure}
\includegraphics[scale=0.25]{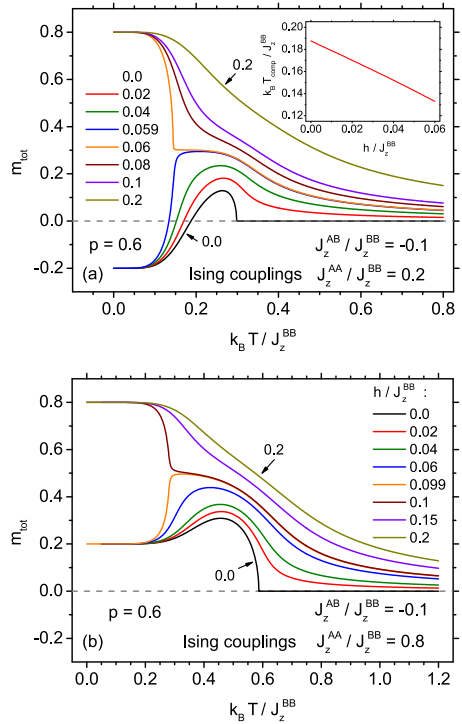}
\caption{\label{fig:fig3} The temperature dependence of total magnetization for the multilayer system in various external magnetic fields. The interaction parameters situate the system (for zero external field) in the ferrimagnetic phase with compensation (a) and without compensation (b). The inset in (a) shows the dependence of the compensation temperature on the normalized external field.}
\end{figure}

\begin{figure}
\includegraphics[scale=0.25]{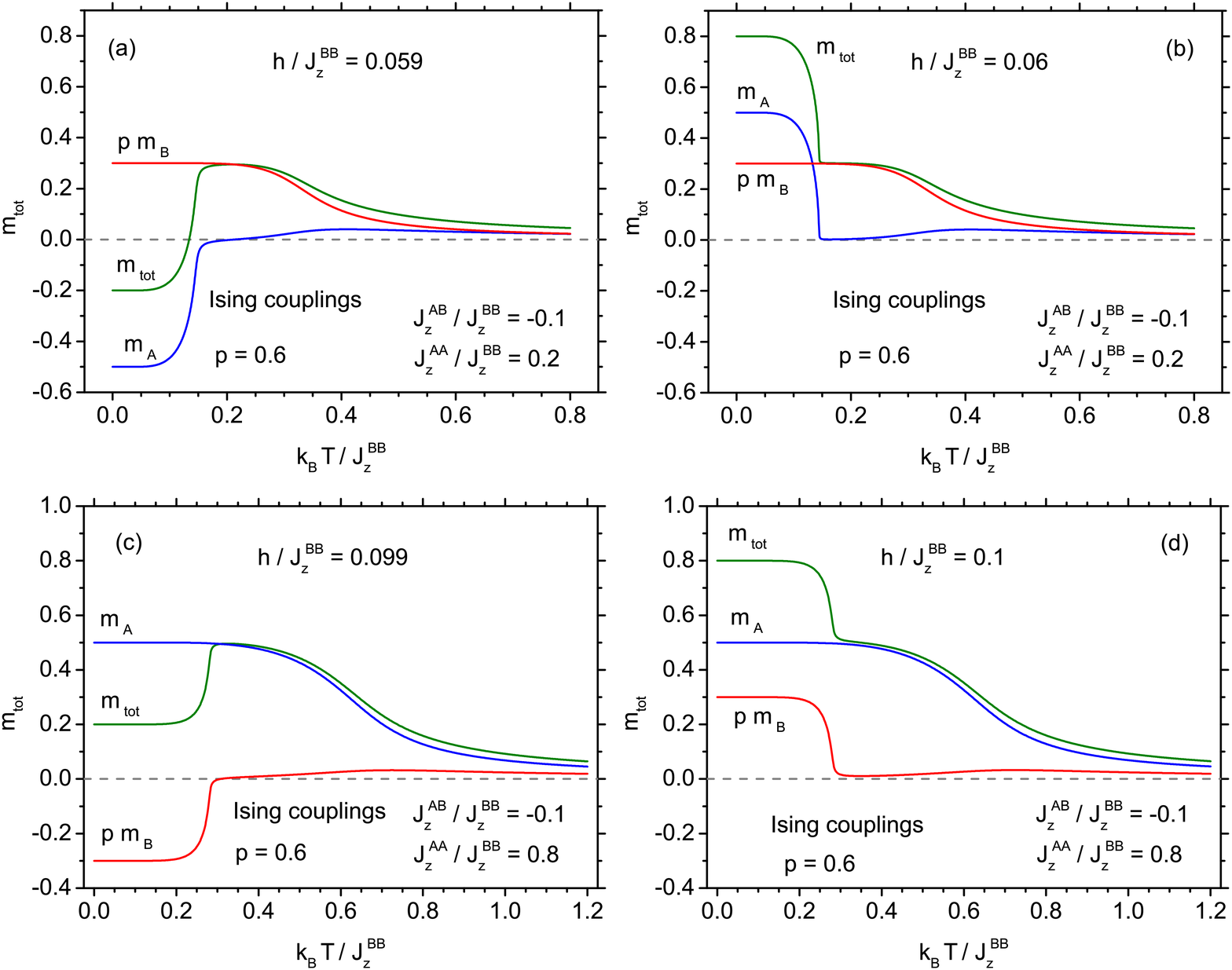}
\caption{\label{fig:fig4}} The temperature dependence of the magnetization of planes A and B as well as total magnetization for the multilayer system. The interaction parameters indicated in the plots locate the system (for zero external field) in the ferrimagnetic phase with compensation (a,b) and without compensation (c,d). The external field in the cases (a,c) is slightly below the critical field for metamagnetic transition, while for (b,d) it is slightly above this critical field.
\end{figure}

At the beginning, it is instructive to determine the range of the model parameters (exchange integrals and concentrations of magnetic atoms) for which a compensation phenomenon is present and is manifested in the temperature dependences of the total system magnetization, at external magnetic field $h=0$. For all couplings of Ising type, this phase diagram is presented in Fig.~\ref{fig:fig2}(a). Each solid line corresponds to some fixed concentration $p$ of magnetic component in plane B. The area of the $J^{AB} - J^{AA}$ plane to the left of each line contains the phase diagram points for which a compensation phenomenon takes place for some temperature. On the contrary, this phenomenon is absent for the phase diagram points to the right of each solid line. It is visible that for large concentrations of magnetic component in plane B, e.g. $p=0.9$, the line separating two phases is strongly inclined and the critical value of $J^{AB}$ is very sensitive to intralayer coupling $J^{AA}$. When concentration $p$ drops, the 
inclination of the phase boundary decreases and the phase with compensation tends to occupy only the upper left corner of the diagram (i.e. the area with weak interlayer couplings $J^{AB}$ and weak intralayer interactions $J^{AA}$). For concentrations lower than $p \lesssim 0.5$ no compensation takes place for all the area of the diagram.

The situation is somehow different if the intraplanar couplings $J^{AA}$ and $J^{BB}$ are both of isotropic Heisenberg type, which is the case illustrated in Fig.~\ref{fig:fig2}(b). There, the boundary separating the phase with and without compensation is almost vertical for high $p$ unless interplanar coupling is very weak (when the boundary tends to be almost horizontal, with small slope). All the phase boundaries meet in the point in which both $J^{BB}$ and $J^{AB}$ vanish. When $p$ decreases, the phase boundary shifts to the lower values of $J^{AA}$. Unlike in the case of all Ising couplings, the phase with compensation survives in the diagram range with weak interactions $J^{AA}$ in principle for all considered values of $J^{AB}$, provided that $p\gtrsim 0.70$. If $p\lesssim 0.70$, only the phase without compensation remains. 

In order to describe the behaviour of the system in question in external magnetic field $h>0$, let us first analyse the dependence of total magnetization on the temperature for two distinct points of the phase diagram, corresponding either to the case with compensation or without compensation. For illustration, the system with all-Ising couplings was selected. Fig.~\ref{fig:fig3}(a) presents the influence of the external magnetic field which, in the ground state, is chosen as opposite to the total magnetization, on the temperature dependence of this magnetization. The system parameters correspond here to occurrence of compensation phenomenon (weak interlayer coupling $J^{AB}$ and weak intralayer coupling $J^{AA}$) for quite low concentration $p$.
 
It should be explained here that in the external magnetic field equal to zero ($h=0$) the system possesses up-down symmetry. This means that there exist two solutions for the total magnetization, which are of the same magnitude but have opposite sign. According to Landau theory of phase transitions, these symmetrical solutions correspond to the same energy, and in the ground state they are separated by the energy maximum. The presence of this separating maximum  opens the possibility to apply a small external magnetic field, oriented  parallel or antiparallel to the total magnetization, without changing  the magnetization direction in the ground state. However, when the field is applied, the energy becomes asymmetric vs. total magnetization, and, for instance, the solution ($m_{tot}<0$) which is opposite to the field ($h>0$) corresponds to a metastable state. In this case, when the field increases and reaches its critical value, the spin-flip transition takes place and the total 
magnetization becomes re-oriented  parallel to the field. After this transition, which is of the 1-st order, the signs of $m_{tot}$ and $h$ are the same, and the system falls into the stable state, where the Gibbs energy is in its absolute minimum. Taking this into account,  we think that such a kind of transition from metastable to stable states becomes interesting for studies in the context of MCE.

To begin with, the temperature behaviour of total magnetization should be calculated. It is known, that temperature itself does not change the symmetry of the Gibbs potential, however, it diminishes the energy barrier between metastable and stable states. Thus, increasing temperature enables the spin-flip transition. In Fig.~\ref{fig:fig3}(a) the dependence $m_{tot}(T)$ for the fields $h/J^{BB}_{z}<0.06$ indicate the compensation at the temperature $T_{comp}$, which shifts to lower values when $h$ increases. The inset in Fig.~\ref{fig:fig3}(a) shows the dependence of the compensation temperature on the external magnetic field and proves the linear type of dependence $T_{comp}(h)
$ up to some critical field $h/J^{BB}_{z}\simeq 0.06$. When $h$ approaches this critical value, the reorientation of the magnetization direction becomes increasingly abrupt. Above the critical magnetic field, the compensation vanishes and magnetization becomes a monotonous, decreasing function of temperature, with a kink in the vicinity of former compensation temperature. However, this kink disappears fast when $h$ further increases.

If the intralayer interaction $J^{AA}$ becomes strong, as in Fig.~\ref{fig:fig3}(b), the qualitative behaviour of magnetization vs. temperature changes. Let us remind that no magnetic compensation is predicted for that range of parameters. Below the critical field the total magnetization first rises and then drops with temperature, but always keeps the same sign. If the critical field $h/J^{BB}\simeq 0.1$ is exceeded, again the magnetization dependence on temperature becomes monotonous and bears similarity to the previous case shown in Fig.~\ref{fig:fig3}(a). 

The behaviour of total magnetization can be explained in details by focusing on magnetizations of both types of magnetic planes, A and B. Their temperature dependences for representative parameters are presented in Fig.~\ref{fig:fig4}. First let us analyse the behaviour of the system in the parameter range for which compensation occurs, just below the critical magnetic field $h/J^{BB}\simeq 0.06$ (Fig.~\ref{fig:fig4}(a)). 
In accordance with Fig.~\ref{fig:fig3}(a) such situation corresponds to the metastable state,
where at low temperature both planes indicate magnetizations oriented antiferromagnetically, and the total magnetization is opposite to the external field. Magnetization of plane A, with weaker intraplanar couplings that plane B, has rectangular-like temperature dependence and abruptly reaches very low values close to the compensation temperature. In the vicinity of this temperature $m_{A}$ varies quasi-linearly and changes its orientation to parallel to $m_{B}$ at the temperature slightly higher than $T_{comp}$. The magnetization of plane B is almost constant in that temperature range and starts to decrease for temperatures significantly higher than compensation temperature. Moreover, it keeps the same orientation for all temperatures.  

For external magnetic field exceeding $h/J^{BB}\simeq 0.06$, illustrated in Fig.~\ref{fig:fig4}(b), the situation changes qualitatively, since no compensation occurs. Starting from the lowest temperatures, both magnetizations, $m_{A}$ and $m_{B}$ indicate parallel orientation. The temperature dependence of $m_{B}$ is very similar to the previous case (Fig.~\ref{fig:fig4}(a)). The magnetization $m_{A}$ again shows rectangular-like thermal dependence for low temperatures. However, after the drop to low values, it remains constant in some range of temperatures in the vicinity of former $T_{comp}$. Then it increases slightly for higher temperatures. Therefore, for the set of parameters allowing for compensation at fields $h$ low enough, the behaviour of the total magnetization with abrupt change at some temperature close to $T_{comp}$ is due to reorientation of the magnetization of plane A, which is characterized by weaker intraplanar couplings than plane B and no dilution.

The situation is different for the set of parameters for which the phenomenon of compensation does not take place at any temperature and magnetic field, what is the content of Fig.~\ref{fig:fig4}(c) and (d). Below the critical magnetic field $h/J^{BB}\simeq 0.1$, it is the diluted plane B which indicates a rectangular-like temperature dependence of magnetization, with a plateau at intermediate temperatures and change of orientation. In principle, the behaviour is analogous to one presented in Fig.~\ref{fig:fig4}(a), but with roles of A and B planes being swapped. Exactly the same comment applies to the case of the stronger external field, exceeding the critical value, shown in Fig.~\ref{fig:fig4}(d). In both figures, the magnetization of undiluted plane has weaker temperature dependence and constant sign. The reorientation transition concerns the plane B with dilution, and since the absolute value of magnetization is reduced with respect to plane A, no change of sign is visible in the total magnetization of 
the system.

\begin{figure}
\includegraphics[scale=0.25]{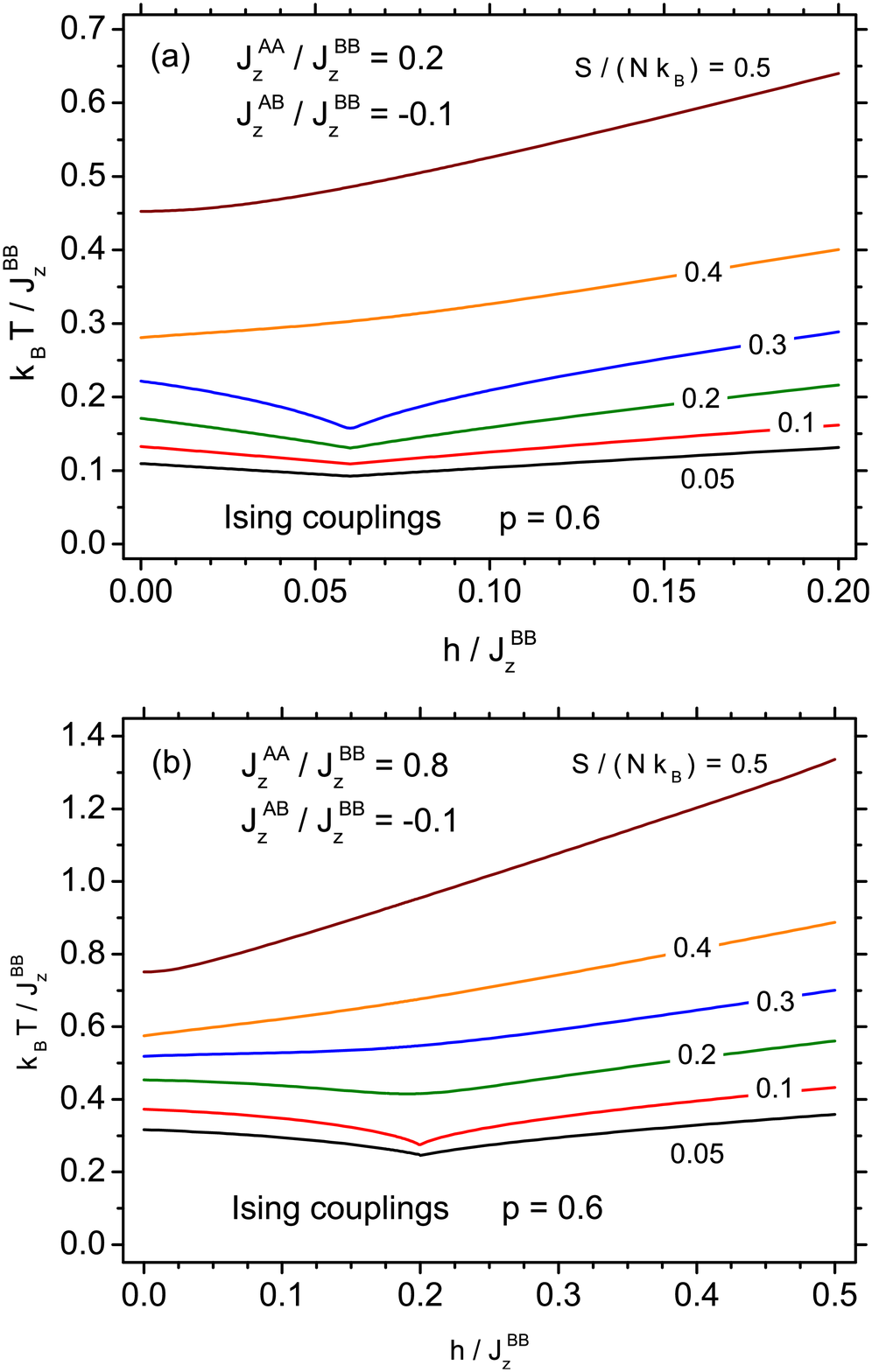}
\caption{\label{fig:fig10}}  The isentropes for a few values of normalized entropy in temperature-magnetic field variables. All couplings are of Ising type. The system for zero external field is in phase with magnetic compensation (a) or without magnetic compensation (b).
\end{figure}

The main quantity the knowledge of which allows to predict the magnetocaloric properties is the magnetic entropy as a function of temperature and external magnetic field. Therefore, let us illustrate its dependence on these parameters in a form of plot of isentropes. Such results are presented in Fig.~\ref{fig:fig10}) for two sets of parameters characterizing the system, i.e. intra- and interaplanar exchange integrals and magnetic component concentration. Namely, Fig.~\ref{fig:fig10}(a) corresponds to the occurrence of magnetic compensation at zero external field, while for Fig.~\ref{fig:fig10}(b) no compensation occurs (the parameters are the same as in Fig.~\ref{fig:fig3}(a) and \ref{fig:fig3}(b), respectively). In both plots it is visible that for high temperature range the entropy always decreases with the increase of the magnetic field. However, for lower temperatures, the field first causes an increase in entropy, and after exceeding some critical field the tendency reverts. The decrease of the magnetic entropy with the magnetic field gives rise to normal MCE, while the increase marks the occurrence of inverse MCE. Therefore, occurrence of inverse MCE can be expected in a considered system in general for low external magnetic fields. The pronounced dip in the isentropes marks the critical field separating both regimes. It should be stated that for the case with possible magnetic compensation, the non-monotonicity of entropy as a function of field extends up to higher entropy values. The behaviour of entropy as a function of external field is fully consistent with results shown in Fig.~\ref{fig:fig3}. Let us emphasize here the thermodynamic identity $\left(\partial S/\partial h\right)_{T}=\left(\partial m/\partial T\right)_{h}$, with the help of which the dependencies $m(T,h)$ and $S(T,h)$ can be analysed jointly. In particular, inverse MCE corresponds to $\left(\partial m/\partial T\right)_{h}<0$, which condition is fulfilled for low magnetic fields both for phase with and without compensation, as seen in Fig.~\ref{fig:fig3}).

\begin{figure}
\includegraphics[scale=0.25]{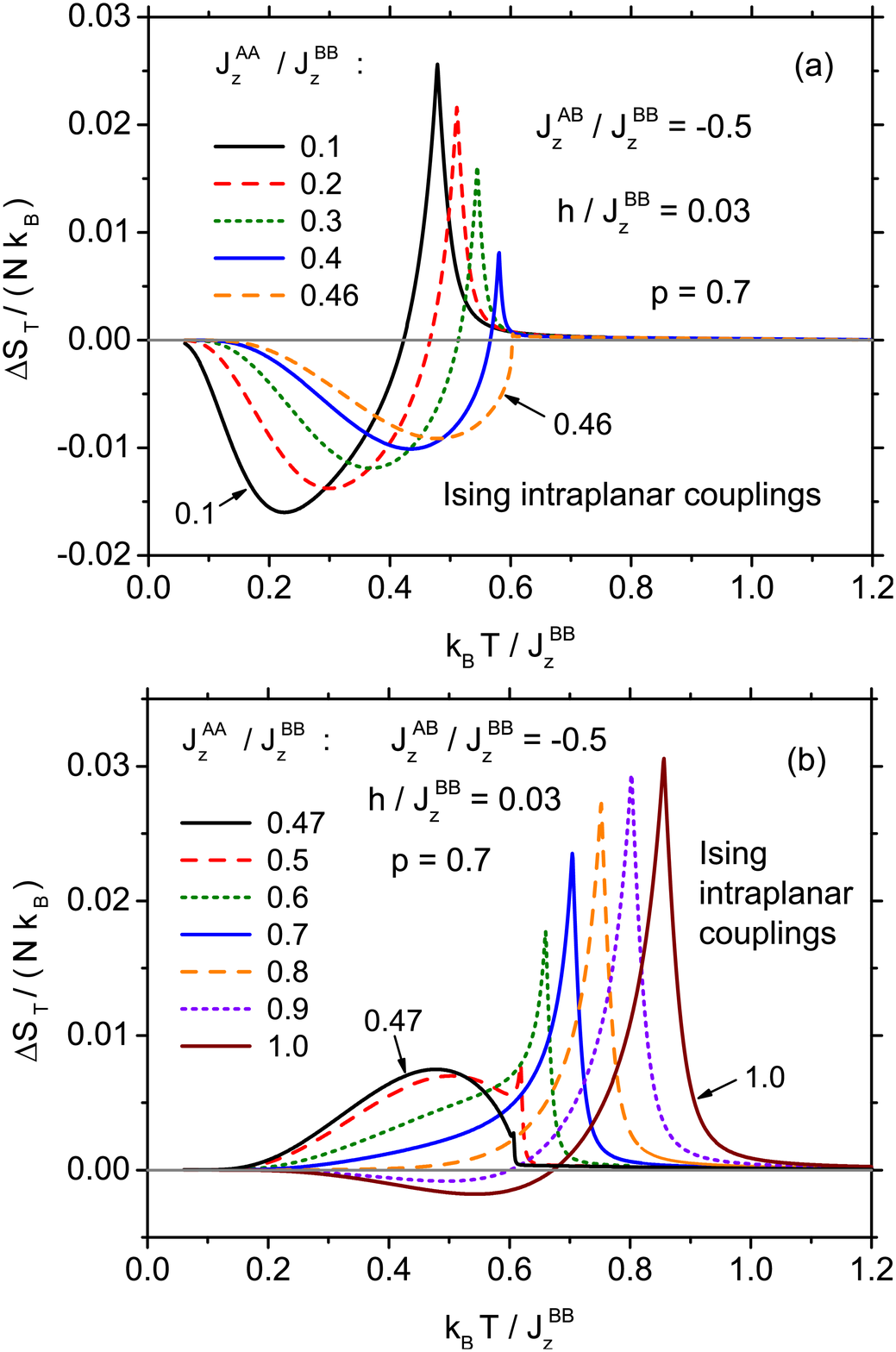}
\caption{\label{fig:fig5}} The normalized isothermal entropy change as a function of normalized temperature for fixed external magnetic field amplitude and varying intraplanar coupling in plane A. All couplings are of Ising type. The system for zero external field is in phase with magnetic compensation (a) or without magnetic compensation (b).
\end{figure}

\begin{figure}
\includegraphics[scale=0.25]{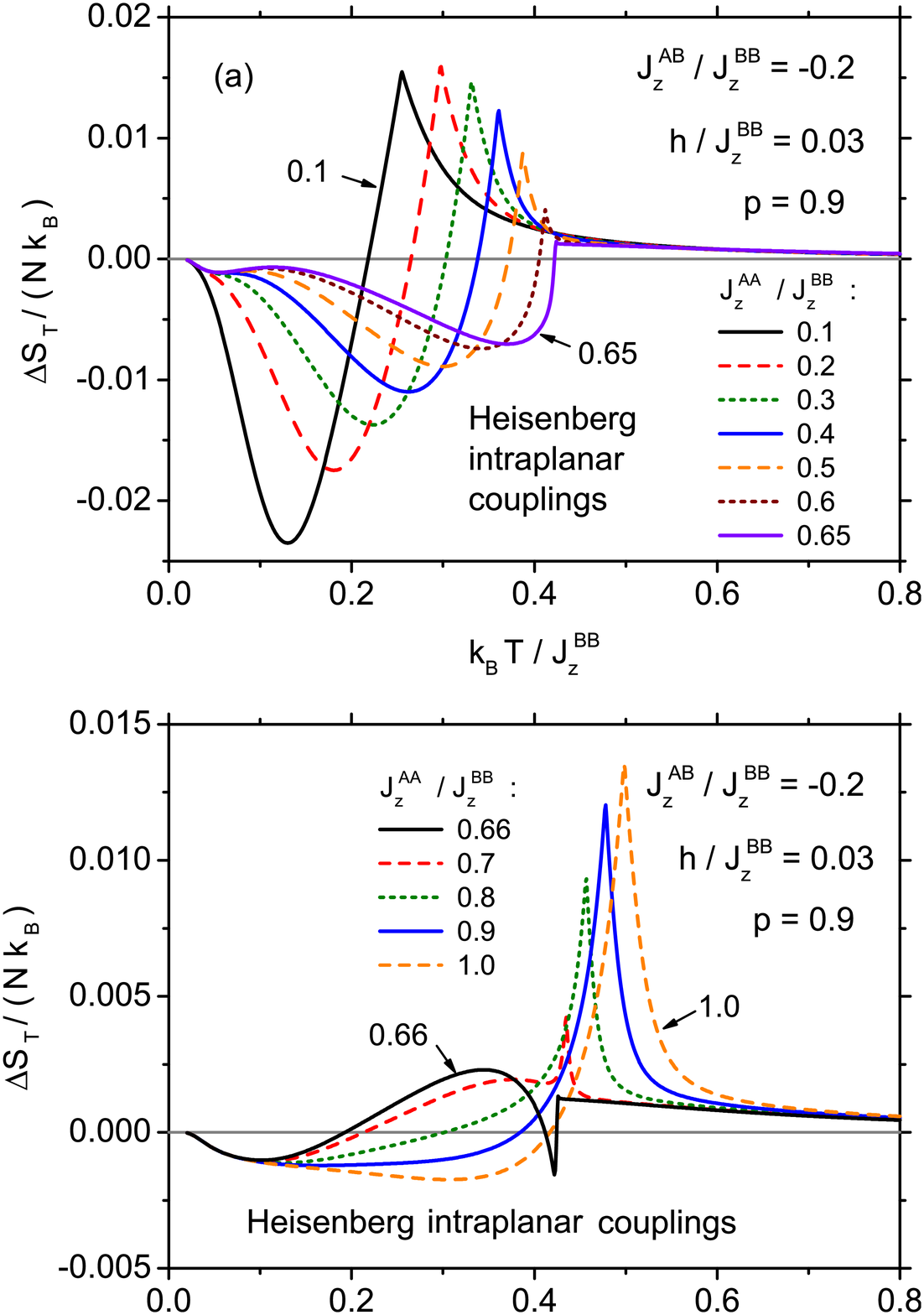}
\caption{\label{fig:fig6}} The normalized isothermal entropy change as a function of normalized temperature for fixed external magnetic field amplitude and varying intraplanar coupling in plane A. All intraplanar couplings are of isotropic Heisenberg type. The system for zero external field is in phase with magnetic compensation (a) or without magnetic compensation (b).
\end{figure}

\begin{figure}
\includegraphics[scale=0.25]{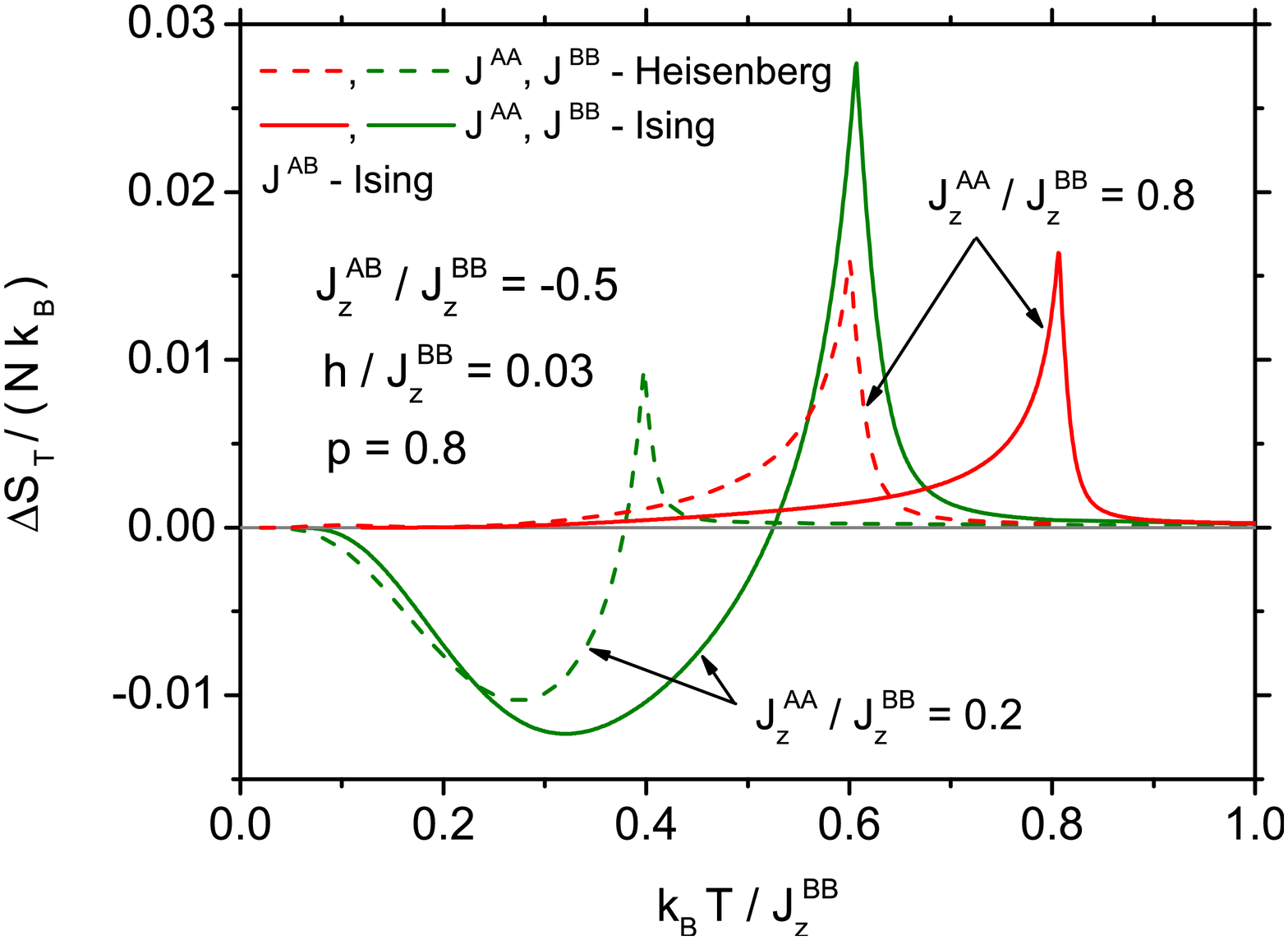}
\caption{\label{fig:fig7}} The normalized isothermal entropy change as a function of normalized temperature for fixed external magnetic field amplitude and varying intraplanar coupling in plane A. All intraplanar couplings are of Ising type (solid lines) or isotropic Heisenberg type (dashed lines). The system for zero external field is in phase with magnetic compensation ($J^{AA}_{z}/J^{BB}_{z}=0.2$) or without magnetic compensation ($J^{AA}_{z}/J^{BB}_{z}=0.8$).
\end{figure}

\begin{figure}
\includegraphics[scale=0.25]{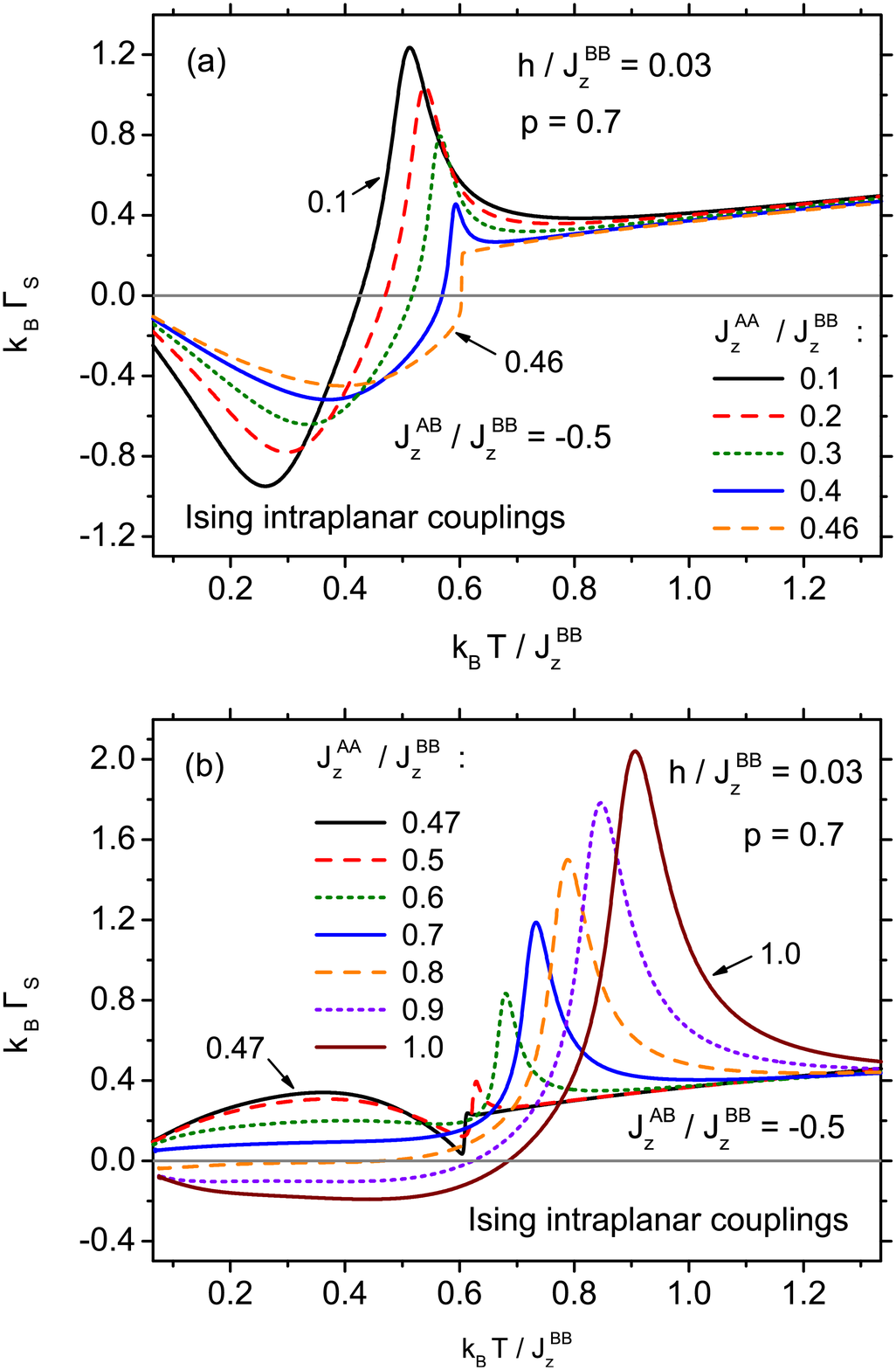}
\caption{\label{fig:fig11}} The normalized isentropic cooling ratio as a function of normalized temperature for fixed external magnetic field and varying intraplanar coupling in plane A. All couplings are of Ising type. The system for zero external field is in phase with magnetic compensation (a) or without magnetic compensation (b).
\end{figure}

\begin{figure}
\includegraphics[scale=0.25]{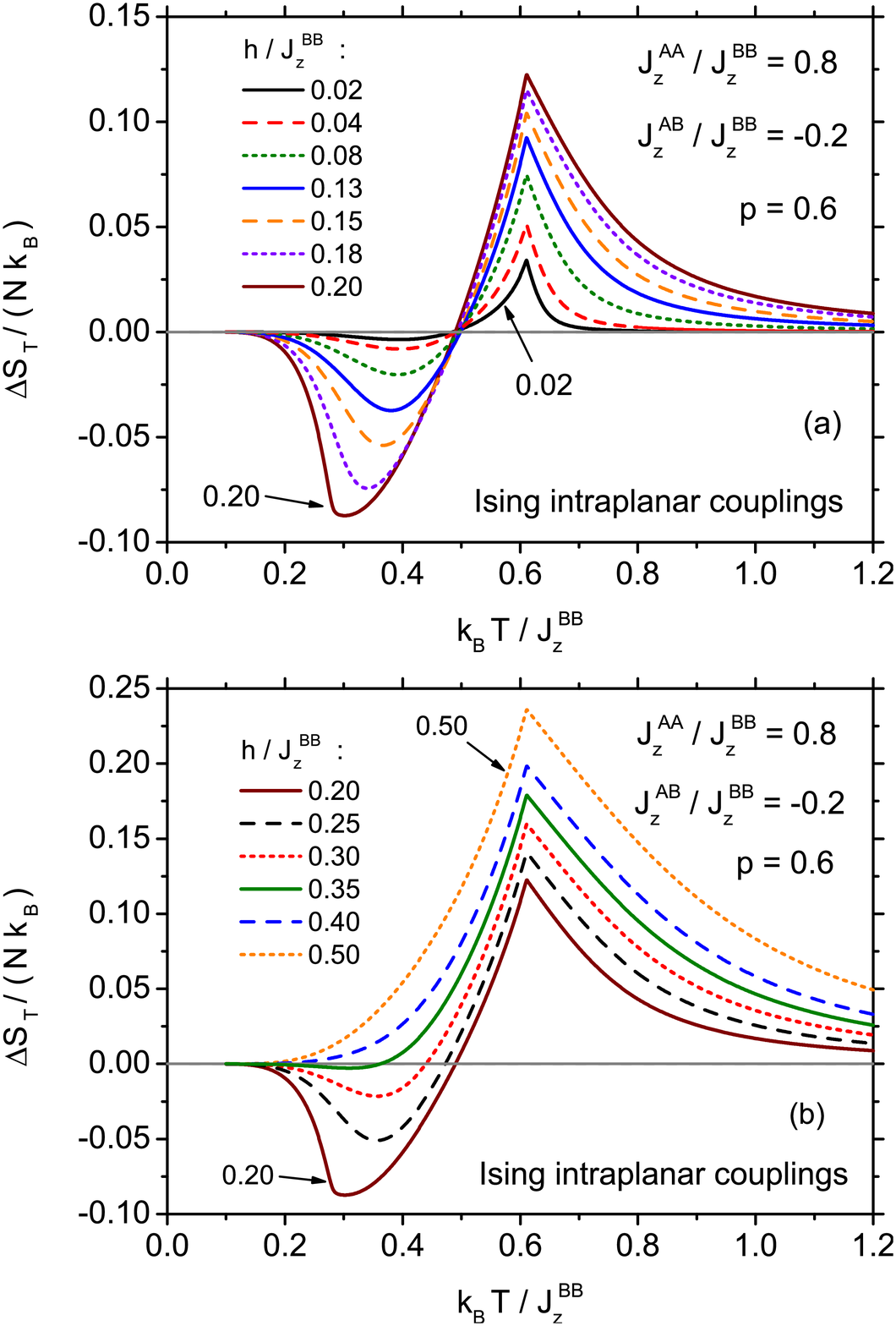}
\caption{\label{fig:fig8}} The normalized isothermal entropy change as a function of normalized temperature for varying external magnetic field amplitude. The system for zero field is in ferrimagnetic phase without compensation. The inverse MCE increases in magnitude for (a) and decreases for (b). All couplings are of Ising type.
\end{figure}

\begin{figure}
\includegraphics[scale=0.25]{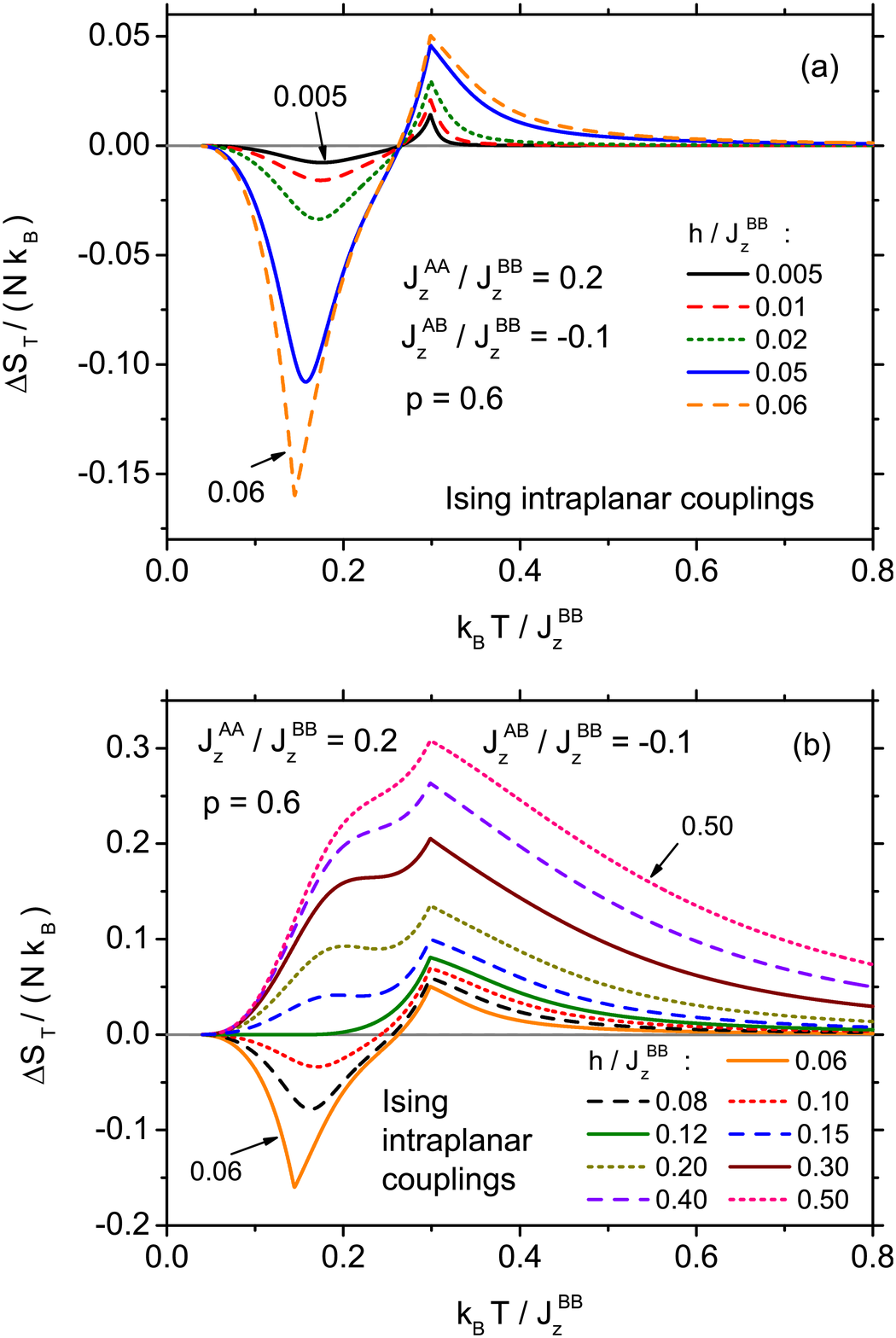}
\caption{\label{fig:fig9}} The normalized isothermal entropy change as a function of normalized temperature for varying external magnetic field amplitude. The system for zero field is in ferrimagnetic phase with compensation. The inverse MCE increases in magnitude for (a) and decreases for (b). All couplings are of Ising type.
\end{figure}

\begin{figure}
\includegraphics[scale=0.25]{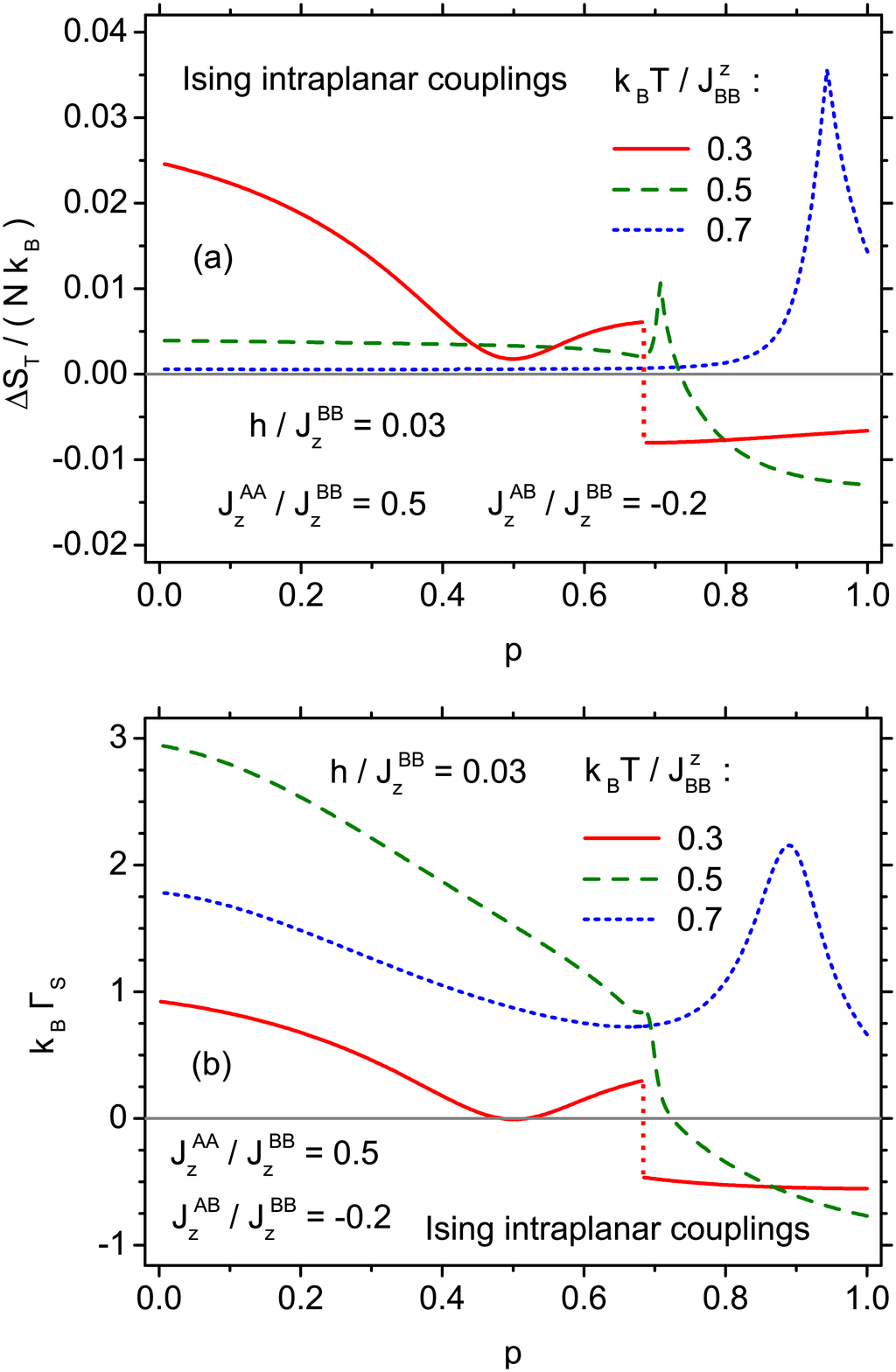}
\caption{\label{fig:fig12}}  (a) The normalized isothermal entropy change and (b) the normalized isentropic cooling ratio as a function of concentration of magnetic component in plane B, for various normalized temperatures. All couplings are of Ising type.
\end{figure}

Having discussed the magnetic entropy vs. temperature and magnetic field, let us focus on the main aim of the study, which is to characterize the magnetocaloric properties of the system. Let us commence from analysis of the magnitude of entropy change when the external field is changed between $h$ and $0$, as a function of the temperature. The results for fixed $h/J^{BB}=0.03$, fixed interplanar coupling $J^{AB}/J^{BB}=-0.5$ and various intraplanar couplings $J^{AA}$ are presented in Fig.~\ref{fig:fig5} (the concentration of magnetic atoms in plane B is $p=0.7$). For the range $0\leq J^{AA}/J^{BB}\lesssim 0.46$ (Fig.~\ref{fig:fig5}(a)), the system is within the regime in which the compensation takes place. It is evident that for lower temperatures, a distinct range of inverse magnetocaloric effect exists, which converts into normal magnetocaloric effect when the temperature is elevated. The minimum in entropy change corresponding to strongest invesre MCE is quite sensitive to coupling $J^{AA}$, since it becomes more shallow and significantly shifts towards higher temperatures when the interplanar 
coupling increases. At the same time, slight shift in the position of the maximum corresponding to normal MCE is also observed, and its magnitude is strongly reduced when $J^{AA}$ tends to the critical value at which the compensation phenomenon vanishes. At that point normal MCE completely disappears and only a pronounced minimum reflecting inverse MCE exists. The situation changes when $J^{AA}$ exceeds the critical value. Just above the critical coupling, it is the inverse MCE that totally disappears, leaving only a broad maximum corresponding to normal MCE exactly at the position of the former negative MCE minimum. Actually the maximum resembles a mirror reflection of that minimum. When intraplanar coupling in plane A increases further, the mentioned maximum tends to vanish and at the same time a peak of normal MCE builds up at higher temperatures. For very strong $J^{AA}$, we deal again with a shallow minimum (inverse MCE) at low temperatures and a pronounced, sharp maximum (normal MCE) at higher 
temperatures.

If the interplanar couplings take isotropic Heisenberg form, the behaviour of MCE is presented in Fig.~\ref{fig:fig6}, for concentration $p=0.9$ and $J^{AB}/J^{BB}=-0.2$. The qualitative features are quite similar to that observed for purely Ising interactions (as in Fig.~\ref{fig:fig5}). Again, if the interaction parameters are such that compensation phenomenon occurs, then a low-temperature range of inverse MCE and high-temperature range of normal MCE is present. If the coupling $J^{AA}$ increases, the high-temperature maximum tends to disappear and inverse MCE minimum shifts towards higher temperatures and becomes less pronounced. After crossing the critical value of $J^{AA}/J^{BB}\simeq 0.65$, for which compensation disappears, the high-temperature peak of normal MCE rebuilds. However, the low-temperature range of inverse MCE does not vanish in the vicinity of critical $J^{AA}$, as it was observed for Ising interactions. Some shallow minimum for lowest temperatures survives and gradually extends towards 
higher temperatures without considerable change in magnitude when $J^{AA}$ rises. Therefore, inverse MCE appears slightly more robust for Heisenberg intraplanar couplings than for all-Ising interactions.

Both cases of intraplanar couplings (Ising or isotropic Heisenberg) are compared for the same interaction parameter $J^{AB}/J^{BB}=-0.5$ and concentration $p=0.8$ in Fig.~\ref{fig:fig7}. The plots present the entropy change as a function of the temperature for weak ($J^{AA}/J^{BB}=0.2$) and strong ($J^{AA}/J^{BB}=0.8$) intraplanar interaction within plane A. The former value corresponds to the case with compensation for both interaction anisotropies, while the latter implies no compensation phenomenon. In general, the qualitative shape of the functions for Ising and isotropic Heisenberg couplings is analogous. The extrema for isotropic coupling are less pronounced and occur for lower temperatures compared to the Ising couplings.

In order to complete the study, we also analyse the behaviour of another magnetocaloric quantity of interest, namely, the adiabatic cooling rate $\Gamma_{S}$. Its temperature dependence is shown in Fig.~\ref{fig:fig11} for various strengths of interplanar antiferromagnetic coupling. Fig.~\ref{fig:fig11}(a) corresponds to the presence of compensation (at zero field), while in Fig.~\ref{fig:fig11}(b) the compensation is absent. The ranges of interaction parameters are the same as in the case of Fig.~\ref{fig:fig5} showing the temperature dependence of isothermal entropy change, while the external magnetic field is set to $h/J^{z}_{BB}=0.03$. The general qualitative behaviour of $\Gamma_{S}$ is quite similar to the behaviour of $\Delta S_{T}$ quantity. For weaker interplanar couplings (case with possible compensation), for low temperatures a pronounced inverse MCE is seen, with the tendency of shifting of the minimum towards higher temperatures with the increase of $J^{AB}$ interaction. On the contrary, maximum corresponding to normal MCE tends to reduce its magnitude completely when interplanar interaction becomes stronger. Close to the boundary between phase with and without compensation (in zero field), the value of cooling ratio remains almost constant (and corresponds to normal MCE) above the critical temperature, whereas below a range of inverse MCE is present.  For the case of absence of compensation, for interplanar couplings strong enough, the inverse MCE is absent in any temperature range. Further increase in $J^{AB}$ causes the maximum to build up again (with strong shift towards higher temperatures). In parallel, a range of inverse MCE is recovered at low temperatures. The described changes mimic the behaviour of $\Delta S_{T}$ to large extent (compare Fig.~\ref{fig:fig5}). 

It is also interesting to follow the evolution of the entropy change magnitude when the amplitude of the external magnetic field $h$ is changed. Such dependencies are presented in Figs.~\ref{fig:fig8} and ~\ref{fig:fig9}. First of them concerns the case of the multilayer with strong intraplanar interactions in plane A, thus it is prepared for the system exhibiting no compensation phenomenon (Fig.~\ref{fig:fig8}(a,b)). For the range of external fields not exceeding the critical field of $h/J^{BB}\simeq 0.20$ (Fig.~\ref{fig:fig8}(a)), a low-temperature, broadened minimum of inverse MCE and a sharp high-temperature maximum of normal MCE occur. When external field $h$ amplitude increases, the magnitudes of both extrema also increase and remain somehow comparable. The high-temperature maximum has quite stable position, while the low-temperature minimum tends to shift to lower temperatures. If the field amplitude exceeds the critical field (Fig.~\ref{fig:fig8}(b)), the situation 
changes. Namely, the inverse MCE range 
tends to decrease its magnitude and then vanishes completely, so that only a single maximum (showing normal MCE) remains. This maximum increases in magnitude monotonously when $h$ rises. Therefore, the inverse MCE is present only for a limited range of external fields $h$. 

If the system is in regime of parameters with compensation, like in Fig.~\ref{fig:fig9}(a,b), the situation is qualitatively similar. However, below the critical magnetic field ($h/J^{BB}\simeq 0.06$ in this case), the inverse MCE dominates over normal MCE when the magnitude of the effect is taken into account, as the low-temperature minimum becomes very deep (Fig.~\ref{fig:fig9}(a)). However, above the critical field (Fig.~\ref{fig:fig9}(b)), this low-temperature range of inverse MCE again tends to vanish completely, while the high-temperature maximum builds up monotonously. For $h$ large enough, a low-temperature minimum transforms into a kind of maximum or kink, the position of which corresponds to increasingly higher temperatures and finally tends to merge with the main maximum. We can conclude that for that choice of interaction parameters (Fig.~\ref{fig:fig9}(a)) the inverse MCE is more pronounced , but occurs in a more narrow range of field amplitudes $h$, whereas for larger fields a 
more broad maximum of normal MCE emerges (Fig.~\ref{fig:fig9}(b)).

In order to illustrate the influence of the magnetic dilution of plane B on the magnetocaloric properties, we show Fig.~\ref{fig:fig12}. In Fig.~\ref{fig:fig12}(a) isothermal entropy change between $h/J^{z}_{BB}=0.03$ and $h/J^{z}_{BB}=0$ is plotted against concentration of magnetic component in plane B for three normalized temperatures, while Fig.~\ref{fig:fig12}(b) shows analogous dependence of normalized isentropic cooling rate $\Gamma_{S}$ at $h/J^{z}_{BB}=0.03$. Let us mention that for zero magnetic field, for $p\lesssim 0.683$ magnetic compensation is absent, while for $p\gtrsim 0.683$ the compensation phenomenon takes place (compare with Fig.~\ref{fig:fig2}(a)). It is visible that for the highest considered temperature, the MCE remains normal in the whole range of concentrations (for both studied magnetocaloric quantities) and achieves a pronounced maximum when this particular temperature becomes the critical temperature of the system. For the lower temperature, the MCE is an inverse one for high concentrations $p$. Dilution causes the effect of switching to a normal one. For the lowest studied temperature, again entropy change and cooling rate are negative for high $p$. When $p$ is reduced, they change sign in a discontinuous way while crossing the boundary between phase without and with compensation. For lower concentrations they reach a minimum and then increase significantly. This plot supports the statement that for high temperatures the MCE in a studied system is a normal one, with a maximum corresponding to the critical temperature. On the other hand, for low temperatures, the effect is inverse for the phase with compensation and can be switched to normal one by crossing the boundary towards the phase where the compensation is absent.

\section{Conclusions}
In the paper, the coexistence of normal and inverse MCE has been analysed for a magnetic multilayer with antiferromagnetic interlayer couplings and selective dilution of one kind of inequivalent magnetic planes. The thermodynamics of the model was described within Pair Approximation method, which is superior to the commonly applied molecular field approximation in characterization of magnetocaloric properties \cite{Szalowski2011}. In particular, it takes into account the interaction anisotropy in spin space, allowing to distinguish between Ising and isotropic Heisenberg couplings, which is beyond the scope of molecular field-based description. Moreover, the method provides a nontrival description of diluted magnets \cite{Balcerzak2009a,Szalowski2012}, including a nonzero critical concentration and nonlinear dependence of the critical temperature on magnetic component concentration. Within the accepted approach to the thermodynamics, the phase diagram and the conditions for occurrence of magnetic compensation 
phenomenon (the concentration of magnetic component as well as inter- and intraplanar couplings) were discussed for the system in question. The presence of normal as well as inverse MCE has been found in isothermal entropy change. It was found that the inverse MCE may be present for the temperatures lower than normal MCE, mainly in the range of parameters for which the compensation is possible. Very close to the boundary between the phase with and without compensation, it is possible to observe only the inverse MCE or only the normal MCE, depending on the side from which the boundary is approached. The existence of inverse MCE is limited by the amplitude of the external magnetic field, since the increase of the field amplitude first enhances and then reduces inverse MCE, finally promoting only the normal MCE.

It should be stressed that for the cases embracing compensation phenomenon we studied only the most interesting case, namely, where the external field enforces the existence of metastable states. These states undergo discontinuous transition to stable solutions above some critical field values. The first-order transitions are accompanied by the pronounced changes in all magnetic properties, and their description was possible on the basis of the expression for the Gibbs energy.  This possibility can be regarded as an advantage of the method. The studies of other possible situations in the external field still should be done, for instance, when exclusively  stable solutions in the ground state are taken into account. Such  studies require further extensive numerical calculations and will be presented elsewhere.

The presented results allow for indicating the range of parameters of the model for which either both effects or just one of them can be observed as a function of the temperature. Moreover, they show how the MCE is sensitive and how it can be controlled by varying magnetic interactions and concentration of magnetic atoms in layered system. This is vital in the context of possible design and optimization of multilayer ferrimagnets to achieve desired magnetocaloric properties. Moreover, the formalism could be extended for example to the case of more than two uniform magnetic subsystems or to systems with long-range interactions.

\begin{acknowledgments}

This work has been supported by the Polish Ministry of Science and Higher Education on a special purpose grant to fund the research and development activities and tasks associated with them, serving the development of young
scientists and doctoral students.

\end{acknowledgments}

\end{document}